\documentclass[a4paper,11pt]{article}
\usepackage{pos}
\usepackage{lineno}
\title{Neutrino-argon cross-section measurements from the MicroBooNE experiment}

\usepackage{fancyhdr}

\author*{Liang Liu}

\onbehalf{on behalf of the MicroBooNE Collaboration}

\affiliation{
  Fermi National Accelerator Laboratory,\\
  Batavia, IL 60510, USA}

\emailAdd{liangliu@fnal.gov}

\abstract{MicroBooNE is a liquid argon time projection chamber (LArTPC) neutrino detector located along the Fermilab Booster Neutrino Beam and 8 degrees off-axis to the Neutrinos at the Main Injector beam. MicroBooNE collected data from both beams accumulating a large neutrino-argon scattering dataset containing hundreds of thousands of events. Understanding neutrino-argon interactions is crucial for the next generation of neutrino oscillation experiments including DUNE. MicroBooNE has developed pioneering methodologies and novel reconstruction tools in order to benchmark models at very high sensitivity across the interaction phase space, including for ultra-rare channels. This proceeding presents an overview of the most recent MicroBooNE neutrino interaction results. These measurements span inclusive, CC0$\pi$, and rare channels including $\Lambda$, $K^+$ and $\eta$ production, providing invaluable datasets for constraining backgrounds and improving the modeling of neutrino scattering critical for the broader LArTPC neutrino physics program.}

\FullConference{
Presented at the 32nd International Symposium on Lepton Photon Interactions at High Energies, Madison, Wisconsin, USA, August 25-29, 2025
}

\begin{document}
\fancypagestyle{firstpage}{
  \fancyhf{}
  \fancyhead[R]{FERMILAB-CONF-25-0791-CSAID}
}
\thispagestyle{firstpage}

\setcounter{page}{0}
\maketitle

\section{Introduction}

The main scientific goals of neutrino oscillation experiments are to search for charge–parity (CP) symmetry violation, determine the neutrino mass hierarchy, and explore the possible existence of sterile neutrinos. CP violation in neutrino oscillations would constitute a new mechanism for breaking CP symmetry, potentially connected to the observed matter–antimatter asymmetry in the Universe~\cite{Fukugita:1986hr,Buchmuller:2005eh,Pascoli:2006ie}. Determining the neutrino mass hierarchy would significantly constrain theoretical models, ruling out roughly half of the current possibilities~\cite{Qian:2015waa}. The existence of one or more species of sterile neutrinos, which do not participate in weak interactions, has been proposed to explain several anomalies reported by previous neutrino experiments~\cite{Giunti:2019aiy}. 
One of the main challenges in neutrino oscillation measurements is the determination of the energy of the initial-state neutrinos. Because neutrinos interact weakly, their energies can be inferred only indirectly from the detected final-state particles on an event-by-event basis. Therefore, the uncertainties in the measurement of neutrino energy arise not only from the detector configuration and response but also from the modeling of neutrino–nucleus cross-sections. The complex nature of nuclear structure, as a quantum many-body system, makes it extremely challenging to calculate neutrino–nucleus cross-sections from first principles. Several semi-classical theoretical models, implemented in neutrino event generators~\cite{Andreopoulos:2009rq,Mosel:2019vhx}, are used to predict the smearing matrix that relates the reconstructed neutrino energies to the true neutrino energies. For experiments employing complex nuclear targets at accelerator energies (hundreds of MeV to tens of GeV), theoretical uncertainties on neutrino-nucleus cross-section models are typically a leading source of systematic errors. Neutrino cross-section data thus serve as a necessary constraint on the theoretical models (and their implementation in event generators) needed to deliver oscillation-based measurements of neutrino properties. Moreover, understanding rare processes such as neutrino-induced strange particle production and $\eta$-meson production is crucial for beyond the Standard Model searches. The experimental results of neutrino–argon interactions presented in this proceeding are highly valuable for future argon-based neutrino oscillation experiments to achieve the precision needed for their ambitious physics goals.

\section{MicroBooNE detector}

MicroBooNE~\cite{MicroBooNE:2016pwy} is a liquid argon time projection chamber (LArTPC) detector located at Fermi National Accelerator Laboratory (Fermilab). It is exposed to neutrino fluxes from both Fermilab beamlines: the Booster Neutrino Beam (BNB) and the Neutrinos at the Main Injector (NuMI) beam. The active volume of MicroBooNE is rectangular, measuring 10.36 m in length along the beam direction, 2.32 m in height, and 2.56 m in width, and contains 85 tonnes of liquid argon. A uniform 273 V/cm drift field was applied across the TPC active volume. The anode readout assembly features three sense wire planes with a 3 mm pitch. 
Thirty-two photomultiplier tubes are located behind the anode plane and covered by acrylic disks coated with a wavelength shifter. 
When charged particles traverse the TPC, the three wire planes record signals induced by the drifting ionization electrons, while the PMT array measures the prompt scintillation light produced during the excitation and ionization of argon atoms, providing precise timing information.

\section{Cross-section of neutrino-argon interactions}

\begin{figure}[htbp]
    \centering
    \includegraphics[width=0.35\linewidth]{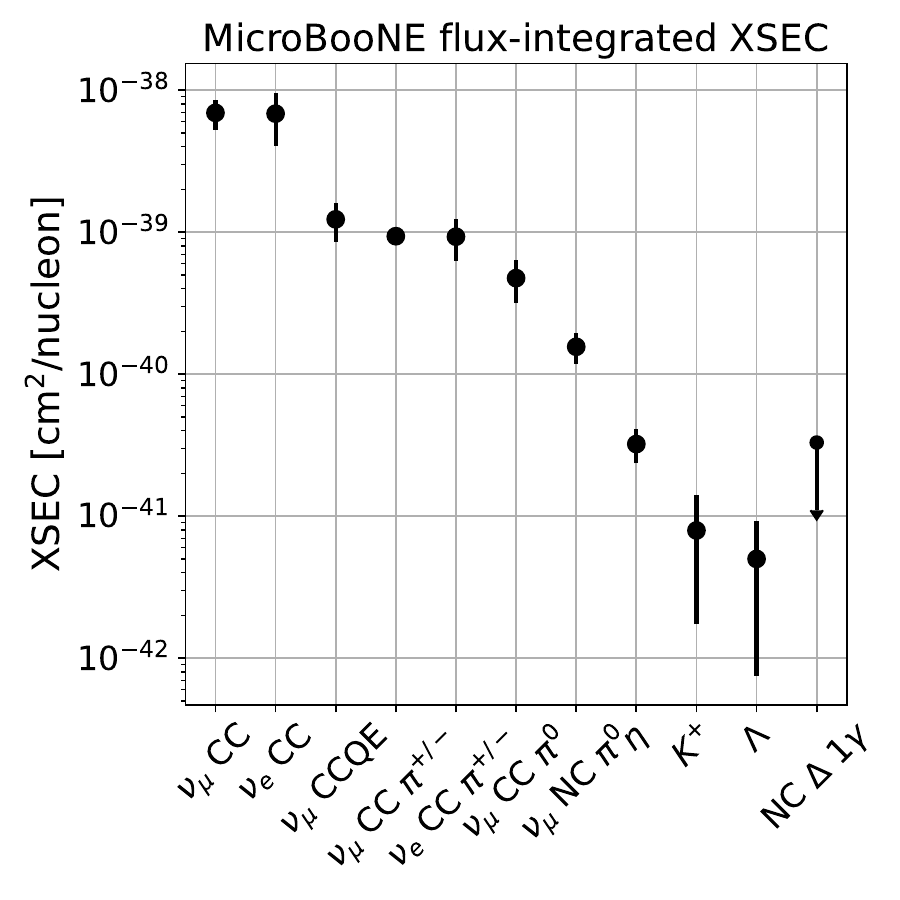}
    \caption{Examples of the flux-integrated cross-section measurements performed by MicroBooNE.}
    \label{fig:xsec-summary}
\end{figure}

Based on an extensive dataset of neutrino–argon interactions collected by the MicroBooNE detector, exposed to the BNB and the NuMI, dozens of cross-section measurements have been performed. These measurements encompass charged-current inclusive neutrino interactions, pion-less processes, pion production, and other rare channels. Figure~\ref{fig:xsec-summary} shows a subset of the flux-integrated total neutrino cross-sections measured by MicroBooNE, spanning multiple orders of magnitude. Our analysis not only achieves a precise measurement of inclusive $\nu_{\mu}$CC interactions, but also observes $K^{+}$ and $\Lambda$ productions with cross-sections at a remarkable level of $\mathscr{O}(10^{-41})$~${\rm cm}^2$/nucleon. These measurements will be crucial input for constraining theory models and physics research in future Short-Baseline Neutrino (SBN) program~\cite{Machado:2019oxb} and Deep Underground Neutrino Experiment (DUNE)~\cite{DUNE:2020jqi}.

\subsection{Cross-section of inclusive $\nu_{\mu}$ charged-current}

A measurement of inclusive muon neutrino charged-current interactions ($\nu_{\mu}$CC) only considers $\mu^{-}$ in the final states, regardless of the particles that emerge from the nucleus. 
In the BNB energy range, quasi-elastic (QE) interactions dominate; however, resonance production (RES), meson-exchange currents (MEC), and deep inelastic scattering (DIS) also make significant contributions. Theoretical models must simultaneously account for multiple scattering mechanisms, in-medium nuclear modifications to the primary neutrino interactions, and final-state interactions (FSI) involving the hadronic reaction products as they exit the nucleus.

\begin{figure}[htbp]

    \centering
    \includegraphics[width=0.75\linewidth]{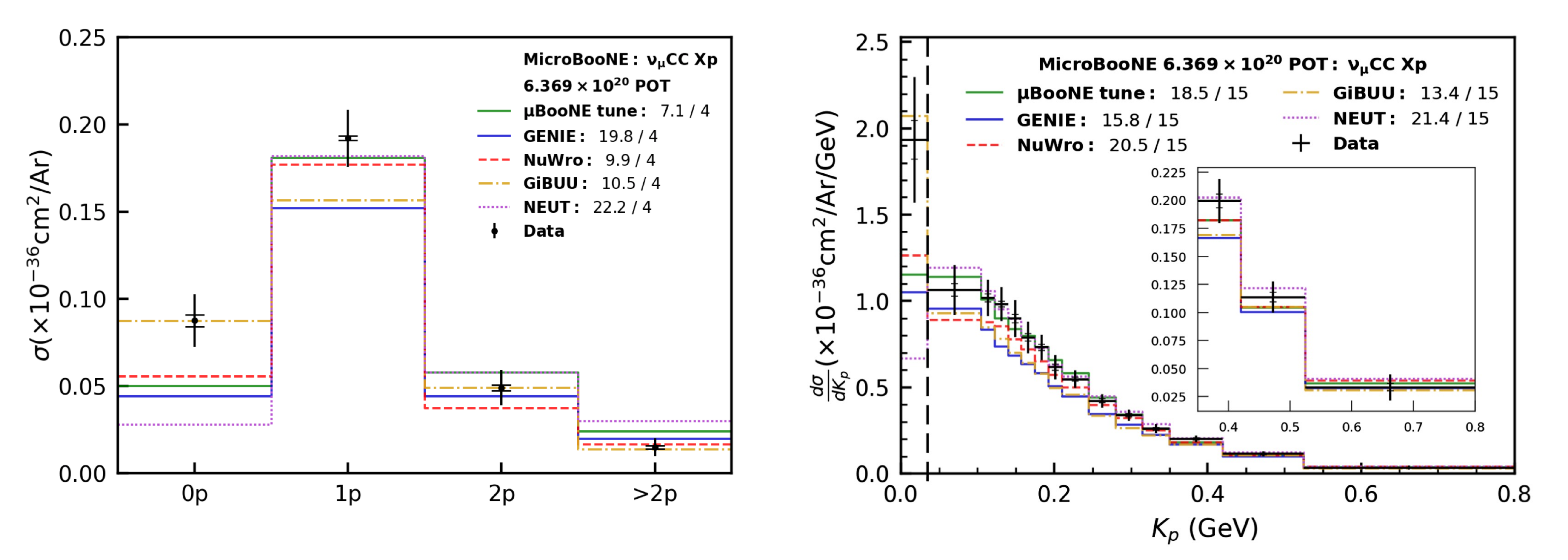}
    \caption{The unfolded differential cross-section for the inclusive $\nu_{\mu}$CC events with zero or multiple reconstructed protons as a function of proton multiplicity (left)~\cite{MicroBooNE:2024zwf} and the leading proton's kinetic energy (right)~\cite{MicroBooNE:2024zkh}. The dashed line indicates the 35 MeV tracking threshold, below which is a single bin that includes events without protons. The unfolded data points show both statistical and systematic uncertainties.}
    \label{fig:proton-multi}
\end{figure}

The MicroBooNE reports the first double-differential cross-section measurements with respect to hadronic final states with zero or multiple reconstructed protons ("0p" and "Np") that have kinetic energy greater than 35 MeV for $\nu_{\mu}$CC~\cite{MicroBooNE:2024zwf, MicroBooNE:2024zkh}. 
Figures~\ref{fig:proton-multi} show measured cross-section with respect to proton multiplicity and the kinetic energy of the leading proton $K_{p}$. The first bin of the $K_{p}$ cross-section includes events from the $\nu_{\mu}$CC 0p channel. 
Only \texttt{GiBUU}~\cite{Buss:2011mx} is able to describe the 0p bin data, which gives it the lowest $\chi^2$ despite its underestimate of high energies. 
These measurements provide valuable new information that will help refine theoretical models and event generators to meet the precision demands of future neutrino oscillation experiments and searches for physics beyond the Standard Model.

\subsection{Cross-section of pion-less $\nu_{\mu}$ charged-current}

\begin{figure}[htbp]
    \centering
    \includegraphics[width=1.0\linewidth]{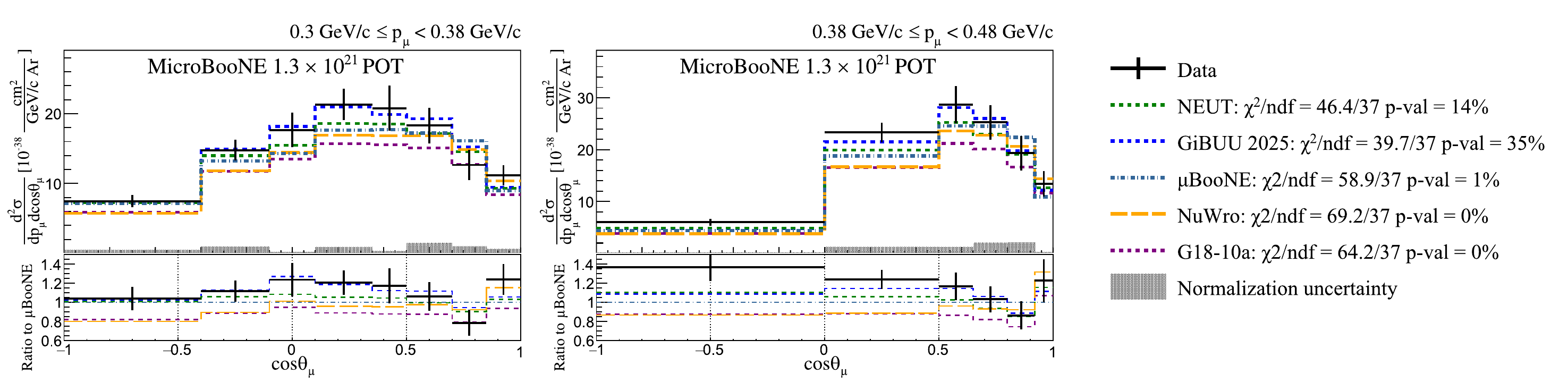}
    \caption{Extracted flux-integrated double-differential cross-sections for the CC$0\pi$ signal with respect to $\cos\theta_{\mu}$ in ranges of $p_{\mu}$. The unfolded data points show both statistical and systematic uncertainties. Only the momentum bins, $0.3~{\rm GeV}/c < p_{\mu} < 0.38~{\rm GeV}/c$ and $0.38~{\rm GeV}/c < p_{\mu} < 0.48~{\rm GeV}/c$, are shown here, with the full results available in ref.~\cite{MicroBooNE:2025ooi}}
    \label{fig:cc0pi}
\end{figure}

A less inclusive interaction topology, the charged-current zero-pion (CC$0\pi$) channel, has also been measured by the MicroBooNE~\cite{MicroBooNE:2025ooi}. 
This measurement of CC$0\pi$ semi-inclusive cross-section aims to inform modeling and facilitate comparison across nuclear targets. This effort further makes it possible to perform correlated multi-target cross-section measurements using Cherenkov detectors in the same neutrino beam, including ANNIE~\cite{ANNIE:2015inw}. Using the full $1.3\times 10^{21}$ POT exposure of the MicroBooNE LArTPC experiment in the Fermilab BNB, a high-statistics measurement of the flux-integrated double-differential cross-section with respect to $\cos\theta_{\mu}$ and $p_{\mu}$ has been performed and shown in Fig.~\ref{fig:cc0pi}.The \texttt{GiBUU} and \texttt{NEUT}~\cite{Hayato:2021heg} models show agreement with the correlated joint distribution with $p$ value $>5\%$, while other models yield a poorer description.

\subsection{Neutrino angle reconstruction using CC$1p0\pi$}

The precision of the reconstructed neutrino direction in LArTPCs is investigated using neutrino interactions that produce a single muon–proton pair without additional detected particles (CC$1p0\pi$).
Good estimations of both the neutrino energy and the incoming neutrino direction are crucial for the study of atmospheric neutrinos.
The direction of the muon–proton system, defined as $\vec{b} = \vec{p}_{\mu} + \vec{p}_{p}$, is proposed as an estimator for the incoming neutrino direction, where $\vec{p}_{\mu}$ and $\vec{p}_{p}$ are muon and proton momentum vectors, respectively. 
The precision of the reconstructed neutrino direction can be evaluated using the angle between $\vec{b}$ and the incoming neutrino direction ($\theta_{\rm vis}$), defined as
\begin{equation}
\theta_{\mathrm{vis}}=\operatorname{acos}\left(\frac{\vec{b} \cdot \hat{z}}{|\vec{b}|}\right),
\end{equation}
where $\hat{z}$ stands for the unit vector along the beam direction of incident neutrinos. 
Nuclear effects have a significant impact on the reconstructed angular orientation.

\begin{figure}[htbp]
    \centering
    \includegraphics[width=0.9\linewidth]{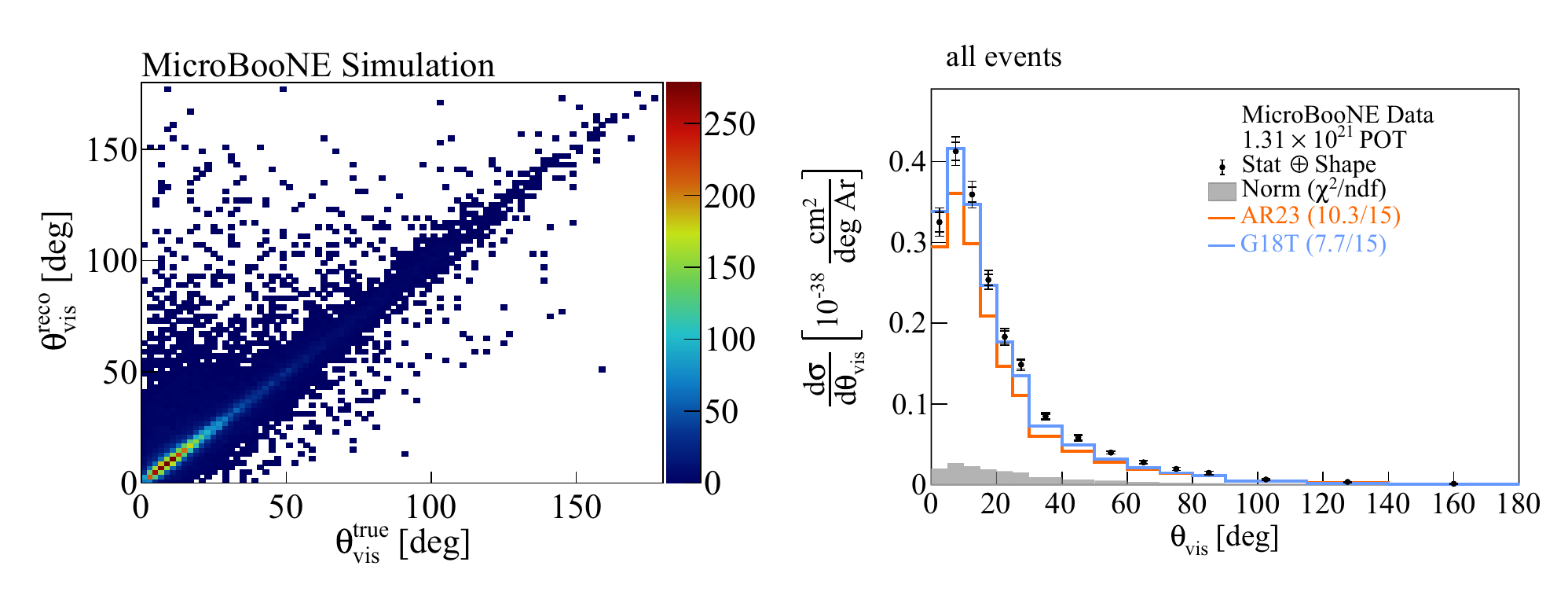}
    \caption{Two-dimensional correlation between the reconstructed and true $\theta_{\rm vis}$ using the selected signal CC$1p0\pi$ simulated events (left); The flux-integrated single-differential cross-sections as a function of $\theta_{\rm vis}$ reported in regularized truth space (right). The unfolded data points show both statistical and systematic uncertainties.}
    \label{fig:neutrino_angle}
\end{figure}

The left plot in Fig.~\ref{fig:neutrino_angle} illustrates the two-dimensional correlation between reconstructed and true $\theta_{\rm vis}$, which  demonstrates that no major biases are observed. For the majority of events, the resolution of $\vec{b}$ is better than $5^\circ$. The right plot in Fig.~\ref{fig:neutrino_angle} shows the single-differential cross-section measurement, while more double-differential cross-section results can be found in ref.~\cite{MicroBooNE:2025phj}. 
This work demonstrates that the neutrino direction reconstruction performance using the single-proton selection is, in most cases, better than using an inclusive selection.
This work demonstrates that current simulations provide a good model of the bias in the inferred neutrino direction for CC$1p0\pi$ events at energies similar to those of atmospheric neutrinos.

\subsection{Pion production}

The DUNE experiment will be exposed to a neutrino flux peaking at energies of a few GeV, where resonance production and deep inelastic scattering are the dominant interaction modes. Recent MicroBooNE measurements of (anti-)electron neutrino charged pion production ($\stackrel{(-)}{\nu}_e$CC${1\pi^{\pm}}$) and neutral-current neutral pion production (NC$\pi^0$) provide essential information for $\nu_{\mu} \to \nu_e$ appearance searches in DUNE.  

\begin{figure}[htbp]
    \centering
    \includegraphics[width=1.0\linewidth]{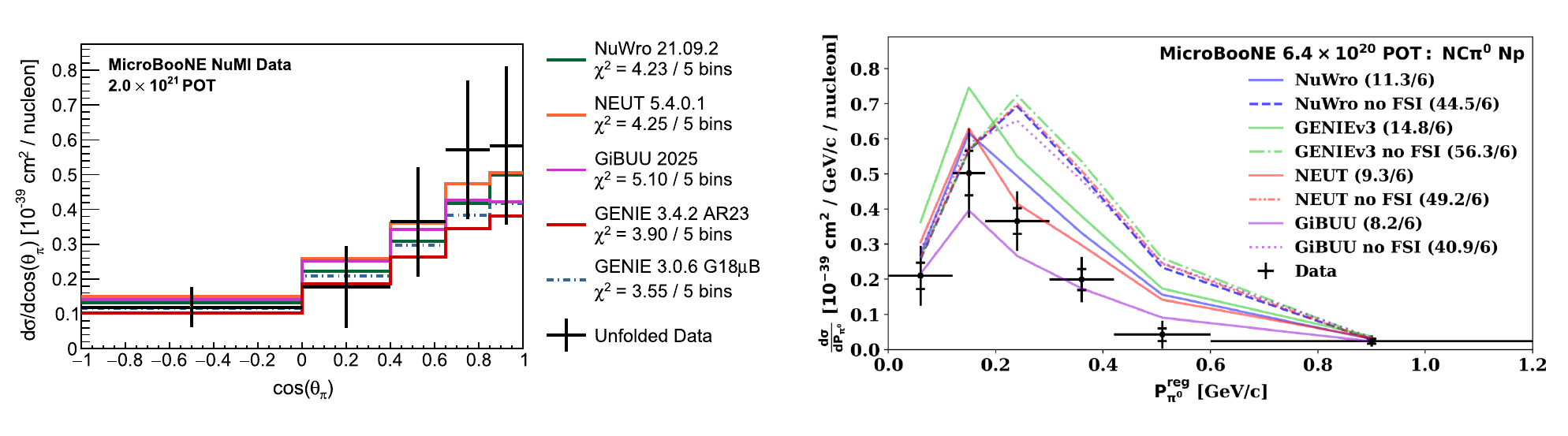}
    \caption{Extracted differential cross-sections in pion angle and pion momentum for $\stackrel{(-)}{\nu}_e \mathrm{CC} \mathrm{Np}$ channel (left) and NC$\pi^0$ channel (right), respectively. } 
    \label{fig:pion_production}
\end{figure}

The first measurement of  $\stackrel{(-)}{\nu}_e$CC${1\pi^{\pm}}$ on an argon target has been performed and the total cross-section is determined to be $(0.93\pm0.13{\rm (stat)}\pm0.27{\rm (syst)})\times10^{-39}$ cm$^2$/nucleon~\cite{MicroBooNE:2025pvb}.
The differential cross-section with respect to pion angle $\cos\theta_{\pi}$ is shown in the left plot of Fig.~\ref{fig:pion_production}. These differential cross-sections are consistent with the predictions from all generators considered. However, the sensitivity of the measurement is currently limited by sizable systematic uncertainties in the flux modeling. 


NC$\pi^0$ production is the primary background in single-shower selections. 
Precise theoretical modeling of NC$\pi^0$ production is thus crucial for these neutrino experiments. 
MicroBooNE reports the first double-differential cross-section measurement of NC$\pi^0$ production in neutrino-argon scattering~\cite{MicroBooNE:2024pdj}. 
The measured differential Np cross-section as a function of $P_{\pi^0}$ is shown in the right plot of Fig.~\ref{fig:pion_production}. The measured single-differential cross section shows that the 0.2-0.5 GeV/$c$ momentum range is strongly impacted by FSIs, suggesting that refinements to FSI modeling may enable a better description of these data. 

\subsection{Rare processes}

Two strange particle production channels, $\bar{\nu}_{\mu} + {\rm Ar} \to \mu^+ + \Lambda + X$ and $\nu_{\mu} + {\rm Ar} \to \mu^- + K^+ +X$, have been measured and their cross-sections are determined to be $2.0_{-1.7}^{+2.2} \times 10^{-40} \mathrm{~cm}^2 / \mathrm{Ar}$~\cite{MicroBooNE:2022cls} and $7.93 \pm 3.27 \text{ (stat.)} \pm 2.92 \text{ (syst.) } \times 10^{-42} \mathrm{~cm}^2 /$ nucleon~\cite{MicroBooNE:2025kqo}, respectively. $\Lambda$ and $K$ production is poorly constrained by existing measurements and is sensitive to the physics of the underlying neutrino interaction and nuclear effects, including nucleon form factors and axial masses, hyperon-nucleus potentials, and final state interactions. Such a process constitutes a potential source of background in proton decay experiments, e.g. $p \to \nu K^+$. A first measurement of $\eta$ meson production in neutrino interactions on argon is also performed and its cross-section is determined to be $3.22 \pm 0.84 \text { (stat.) } \pm 0.86 \text { (syst.) }  \times 10^{-41} {\rm cm^2/nucleon}$~\cite{MicroBooNE:2023ubu}. Higher order resonances such as $N(1535)$, $N(1650)$, and $N(1710)$ decay to an $\eta$ meson with sizable branching ratios. Measuring $\eta$ production in neutrino interactions is a promising way to study RES interactions involving nucleon resonances heavier than $\Delta$ baryons. In addition to the important impact on cross-section modeling, the ability to observe $\eta$ decays in a LArTPC provides a novel tool for the calibration of the electromagnetic (EM) energy scale. 

\section{Summary}

MicroBooNE has reported a broad program of neutrino–argon cross-section measurements, leveraging LArTPC technology to examine scattering physics in unprecedented detail. The topics of MicroBooNE’s measurements range from multi-differential cross-sections of inclusive channels to pion production and rare processes such as strange-particle and $\eta$ meson production.
These results, together with ongoing analyses, will inform precision neutrino-oscillation studies in DUNE and other next-generation experiments.

\bibliographystyle{unsrt}  
\bibliography{references}  

\end{document}